\documentclass{aa}
\usepackage{epsfig}
\usepackage{rotating}
\usepackage{natbib}
\usepackage{graphicx}
\newcommand{\rosn}{{\sl R\"ontgen Satellite}}
\newcommand{\ros}{{\em ROSAT}}
\newcommand{\chan}{{\em Chandra}}
\newcommand{\xmm}{{\em XMM-Newton}}
\newcommand{\vltn}{{\em Very Large Telescope}}
\newcommand{\vlt}{{\em VLT}}
\newcommand{\hstn}{{\em Hubble  Space Telescope}}
\newcommand{\hst}{{\em HST}}
\newcommand{\nttn}{{\em New Technology Telescope}}
\newcommand{\ntt}{{\em NTT}}

\newcommand{\fors}{{\em FORS1}}
\newcommand{\forsn}{{\em FOcal Reducer/low dispersion Spectrograph}}
\newcommand{\forstwo}{{\em FORS2}}
\newcommand{\gsc}{{\em GSC-2}}

\newcommand{\midas}{{\em MIDAS}}

\def \oneight{RX J1856.5$-$3754}
\def \zeroseven{RX J0720.4$-$3125}
\def \zerofour{RX J0420.0$-$5022}
\def \onethree{RX J1308.6+2127}
\def \onesix{RX J1605.3+3249}

\def \oneseven{1RXS\, J214303.7+065419}

\begin{document}

\title{VLT optical observations of the isolated neutron star \\ \zerofour \thanks{Based on observations collected at ESO, Paranal, under Programmes 66.D-0128(A), 078.D-0162(A)}}

\author{R. P. Mignani\inst{1}
\and
C. Motch\inst{2}
\and
F. Haberl\inst{3}
\and
S. Zane \inst{1}
\and
R. Turolla\inst{4,1}
\and
A. Schwope\inst{5}
}

   \institute{Mullard Space Science Laboratory, University College London, Holmbury St. Mary, Dorking, Surrey, RH5 6NT, UK
\and
CNRS, Universit\'e de Strasbourg, Observatoire Astronomique, 11 rue de l'Universit\'e, 67000 Strasbourg, France 
\and 
Max Planck Institut f\"ur Extraterrestrische Physik, Giessenbachstrasse, D85748, Garching, Germany
\and
Department of Physics, University of Padua, via Marzolo 8,  Padua, 35131, Italy 
\and
Astrophysikalisches Institut Potsdam, An der Stenwarte 16, D14482, Potsdam, Germany
}

\titlerunning{Optical observations of \zerofour}

\authorrunning{Mignani et al.}
\offprints{R. P. Mignani; rm2@mssl.ucl.ac.uk}

\date{Received ...; accepted ...}
\abstract{X-ray  observations performed  with the  \rosn\  (\ros) led to the discovery of  seven  radio-silent isolated neutron stars (INSs) which  are detected only through the relatively  dim and  purely  thermal  X-ray  emission from  the cooling star surface. A few  of these INSs (a.k.a.  X-ray Dim INSs, or  XDINSs)  have been also detected  at  optical  wavelengths where  they seem to feature  thermal spectra. Optical studies  of XDINSs thus play a crucial role in  mapping the temperature distribution on the neutron star surface and in investigating the existence of an atmosphere around the neutron star. }{The aim  of this work is to investigate the optical identification  of the XDINS \zerofour,  tentatively proposed  by Haberl et al.  (2004) based on  \vltn\ (\vlt) observations.}{We re-analysed the  original observations of Haberl et al. (2004) to  assess the detection significance of the proposed counterpart and we performed  deeper \vlt\ observations aiming at a higher confidence detection.}{With a  $\sim 2 \sigma$  detection significance and a re-computed flux of $B=27.52 \pm 0.61$, we can not rule out that  the proposed counterpart was spurious and produced by the halo of a very bright nearby star.  While we could not detect the proposed counterpart in our deeper \vlt\ observations,  we found evidence for a marginally significant ($\sim 3.9  \sigma$) detection of a similarly faint object ($B= 27.5\pm 0.3$), $\approx 0\farcs5$ north of it and coincident with the updated \chan\ position of \zerofour. Interestingly, the angular separation is consistent with the upper limit on the \zerofour\ proper motion (Motch et al. 2009), which suggests that we might have actually detected the Haberl et al. proposed counterpart. From the flux of the putative \zerofour\ counterpart we can rule out a $> 7$ optical excess with respect to the extrapolation of the \xmm\ spectrum.}{High spatial resolution observations with the refurbished \hstn\ (\hst) are the only way to confirm the detection of the putative candidate counterpart and to validate its identification with \zerofour.}
 
   \maketitle

\section{Introduction}

X-ray observations performed with the \rosn\ (\ros) yielded to the identification of a group of seven  radio-silent (Kondriatev et  al.  2008)\footnote{The claimed low-frequency pulsed  emission from two of them  (Malofeev et al. 2007) has not been  confirmed yet.} Isolated Neutron  Stars (INSs). Their relatively  dim X-ray emission  ($L_X\approx 10^{30}$--$10^{31}$ erg$\,{\rm s}^{-1}$) originally earned  them the nickname of X-ray Dim INSs,  or XDINSs  (see Haberl~2007;  van Kerkwijk  \& Kaplan  2007, and Kaplan  2008 for recent  reviews).  Recently, a new XDINS candidate has been identified in archival \xmm\ observations (Pires et al. 2009). XDINSs  have purely  thermal X-ray spectra which are best represented by a blackbody ($kT\approx 50$--100 eV), as  expected for middle-aged  ($\sim 1$ Myr) cooling  INSs, whose emission radius is  consistent with a sizable fraction  of the neutron star surface.  The derived hydrogen column densities $N_{\rm H} \approx 10^{20}$ cm$^{-2}$ suggest distances $<$ 500 pc (Posselt et al. 2007), as confirmed in two cases by their optical parallaxes (e.g. van Kerkwijk \& Kaplan 2007). X-ray pulsations  ($P=3$--12 s) have been detected for all of them (Haberl et al.  1997, 1999;
Haberl \& Zavlin 2002;
Hambaryan et al.  2002;
Zane et al.  2005;
Tiengo \& Mereghetti 2007) but \onesix, although with different pulsed fractions. The measurement of the period derivative $\dot{P}$ (Cropper et al. 2004; Kaplan \& van Kerkwijk 2005a; Kaplan \& van Kerkwijk 2005b; van Kerkwijk \& Kaplan 2008) yielded spin-down ages of $\sim 1.5-3.8$ Myrs and rotational energy losses $\dot  {E} \sim (3-5)  \times 10^{30}$  erg s$^{-1}$. Broad  absorption features  ($E_{\mathrm{line}}\approx  0.2$--0.7 keV) have been observed  in all XDINSs but \oneight\  (Haberl et al.  2003, 2004; van Kerkvijk et al.   2004; Zane et al.  2005), superimposed to the  thermal  continuum.  These  features  are  likely  due to  proton cyclotron and/or bound-free, bound-bound  transitions in H, H-like and He-like atoms.  The inferred magnetic fields of $\sim 10^{13}-10^{14}$ ~G  are  consistent  with  the  values  derived  from  the neutron star spin  down   and suggest that  XDINSs might be (evolutionary) linked  to other class  of  INSs, the magnetar candidates (see Mereghetti 2008 for a recent review) and
the Rotating Radio Transients (e.g. Popov, Turolla, Possenti 2006). \\
In the optical, only \oneight\ (Walter \& Matthews 1997; Walter 2001), \zeroseven\ (Motch  \& Haberl  1998; Kulkarni \& van Kerkwijk; 1998; Motch et  al. 2003)  and \onesix\ (Kaplan et al.  2003; Motch et al.  2005; Zane et al. 2006) have counterparts certified by their proper motion measurements, while likely  candidates have been proposed for  \onethree\ (Kaplan et al.  2002) and \oneseven\ (Zane  et al.  2008; Schwope et al. 2009) based on their coincidence with the X-ray positions.  Apart from providing a clear evidence of the optical identification, proper motion measurements are important to obtain an estimate of the kinematic age of the neutron star, to be compared with the characteristic age derived from the spin-down. The XDINS optical fluxes usually exceed by a factor  of $\sim 5$ (or more) the  extrapolation of the X-ray   blackbody, and their optical spectra, when measured, seem to follow a Rayleigh-Jeans  distribution (e.g., Kaplan 2008).   The XDINS optical emission has been interpreted
either in terms of a non-homegeneous surface temperature distribution, with the cooler part emitting the optical  (e.g., Pons et al. 2002),  or of reprocessing of the surface radiation by a thin H atmosphere around a bare neutron star  (Zane et al. 2004; Ho 2007), or of non-thermal emission from particles in the star magnetosphere (Motch et al. 2003).  However, for the  measured $\dot{E}$, magnetospheric emission  would not be detectable, at least if  an average optical emission  efficiency of rotation--powered neutron stars (e.g.  Zharikov et  al. 2006) is assumed.  Alternatively, like for the magnetars, optical magnetospheric emission might be powered by the neutron  star magnetic field, as proposed for \oneseven\ (Zane et al. 2008). \\
One  of  the  XDINSs   without  a  certified  optical  counterpart  is \zerofour.  The first optical observations of the field performed with the \nttn\ (\ntt) soon after the discovery of the X-ray source (Haberl et al.  1999)  did not reveal any candidate  counterpart brighter than B$\sim25.2$  and R$\sim25.2$.   More recently,  thanks to  the updated \chan\ position, a possible optical identification was proposed by  Haberl et  al.  (2004) with  a faint object  (B$=26.6 \pm 0.3$,  V$\ge  25.5$) tentatively  detected  on  archival \vltn\  (\vlt) images.   However,  the  identification  has  not  been  confirmed  so far. The  field of \zerofour\  was also observed in  the near-infrared (NIR) with the \vlt\ but no candidate counterpart was detected down to H$\sim 21.7$ (Mignani et al. 2007; Lo  Curto et al.  2007; Posselt et al. 2009) and K$_s  \sim 21.5$ (Mignani et al. 2008). \\
In  this  paper  we  re-analyze  the original  \vlt\  observations  of \zerofour\  presented  by Haberl  et  al.   (2004)  and we  report  on follow-up, longer  optical observations of  the candidate counterpart, performed by our  team with the \vlt.  Observations  and data analysis are described in Sect. 2, while results are presented and discussed in Sect. 3 and Sect. 4, respectively.

\section{Observations}

\subsection{Observation description}

Optical observations of \zerofour\ were performed in service mode with the  \vlt\ at  the ESO  Paranal observatory  on November  21st  2000, on November 25th 2006, January 16th and 22nd, and  February 11th 2007 (see Tab.\ 1  for a  summary). 
The  2000   observations  were  performed  with   \forsn\  (\fors),  a multi-mode camera for imaging and long-slit/multi-object spectroscopy, as part  of the ESO  guaranteed time programme.   At the epoch  of the
observations  \fors\   was  equipped  with  the   original  four  port 2048$\times$2084 CCD  detector and  it was mounted  at the  \vlt\ Antu telescope.   The observations  were performed  in  standard resolution mode, with  a 0\farcs2 pixel  size and a  field of view  of 6$\farcm8 \times 6\farcm8$.  The low gain,  fast read-out, single port mode was chosen. A sequence of three  1200 s exposures was obtained through the Bessel B  filter, with an airmass of  $\sim 1.14$,  an  image quality of $\sim 0\farcs7$,  and dark  time conditions.  Since the seeing values measured by the  differential image motion monitor (DIMM) are relative to the zenith and not to the  pointing direction of the telescope they are not necessarily indicative of the actual image quality.  We thus computed  the actual image quality from the measured  point spread function (PSF), derived  by fitting the full width half maximum (FWHM) of a number of  well-suited field stars using the {\em Sextractor} tool (Bertin \& Arnouts 1996), as documented in the \fors\ data quality control pages\footnote{http://www.eso.org/observing/dfo/quality/FORS1/qc/qc1.html}.   \\
Two 600 s additional exposures  were obtained  in the V filter but since the proposed candidate counterpart was not detected we focus our analysis  on the B-band data  only.  Sky conditions were  reported to be  photometric (see Haberl et  al.  2004 for a  more detailed observations  description).  A very  bright star, CD-50 1353  ($B=9.9$, as listed in {\em Simbad}), located at $\approx 45\arcsec$  from the position of \zerofour\  was partially masked  using the \fors\  occulting bars. Bias,  twilight flat--fields frames,  and  images of  the standard  star fields SA 92 and Rubin 149 (Landolt 1992) were obtained as part of the \fors\ science calibration plan.

\begin{table}
\begin{center}
  \caption{Log of  the \vlt\ \fors\ and \forstwo\ B-band observations of  \zerofour). Columns report the observing date (yyyy-mm-dd), the number of exposures (N) and the total integration time per night (T), the  image quality (IQ) and rms (in parentheses),  as computed on the image,  and the airmass. Values are the average computed over the exposure sequence. }
\begin{tabular}{lccccc} \\ \hline
& Date & N &  T (s) & IQ  ($\arcsec$)  & Airmass	\\ \hline 
FORS1 & 2000-11-21 &  3 & 3600 & 0.74 (0.08)&  1.14    \\ \hline 
FORS2 &2006-11-25 &  5 & 2915 & 0.71 (0.10)&  1.24  \\ 
&2007-01-16 & 10 & 5830 & 0.93 (0.18)&  1.18  \\ 
&2007-01-22 &  5 &2915 & 0.82 (0.11)&  1.19  \\           
&2007-02-11 &  5 & 2915 & 0.82 (0.12)&  1.30  \\                           \hline 
\end{tabular}
\label{data}
\end{center}
\end{table}

The 2006/2007 observations were performed with \forstwo\ as part of the ESO open time programme.  At the  epoch of the observations, \forstwo\ had swapped with \fors\ at the \vlt\ Antu telescope. \forstwo\ is equipped with  two 2k$\times$4k  MIT  CCD detectors.   Due  to vignetting,  the effective  sky coverage  of  the  two detectors  is  smaller than  the projected detector field  of view, and it is larger  for the upper CCD chip.   Observations were performed  in high  resolution mode,  with a 2$\times$2 binning  and a pixel  size of 0$\farcs$125.  The  low gain, fast  read-out mode  was chosen.   The telescope  pointing was  set in order to position \zerofour\ in the upper CCD chip to include a larger number of reference stars for a precise image astrometry thanks to its larger effective sky coverage ($3\farcm5 \times 2\arcmin$).  Sequences of 580  s exposures  were obtained through  the Bessel B filter.  The bright star CD-50  1353 was more efficiently masked both by positioning it  at the centre of the  gap  between the  two  chips  and  by  using the \forstwo\ occulting  bars. Unfortunately, the distance of \zerofour\ from the gap ($\sim 17\arcsec$) and the width of the occulting bars ($\sim 25\arcsec$), which can move along  one CCD direction only, made it impossible to mask completely the star halo. Exposures  were taken  in dark  time and  under mostly  clear  but not perfectly photometric sky  conditions.  In particular, the night  of February 11th was  affected by  the  presence of  thin  variable cirri.  Atmospheric conditions  were  not  optimal  either.   The first  two  nights  were affected  by a  strong wind,  close to  the telescope  pointing limit, while the nights of January 16th  and February 11th were affected by 40 \% humidity. 
Unfortunately,  although foreseen by the instrument science calibration plan, only for some nights both day and night time  calibration frames were taken.  In  particular, no twilight flat--fields were taken for the night of November 25th, while B-band  standard  star images (of  the  Rubin  152 field)  were taken on  the  night  of   February  11th  only.

\subsection{Data reduction and calibration}

We    retrieved    the   \fors\    science images    from    the   public    ESO  archive\footnote{http://archive.eso.org}  and  we  reduced them  using tools  available  in \midas\  for  bias  subtraction, and  flat--field correction.  The  same reduction steps  were applied to  the \forstwo\ science images  through  the   ESO  \forstwo\  data  reduction pipeline\footnote{http://www.eso.org/observing/dfo/quality/FORS2/pipeline}.   We searched the archive for suitable twilight flat--fields to reduce the  \forstwo\  November  25th science images  but the  closest in time were those associated with our January 16th images.  We evaluated the possibility  of using  lamp  flat--fields as  backup calibration frames, with the caveat that they are affected by reflections produced by the instrument atmospheric dispersion correctors.    However, since lamp flat--fields  are only taken for trending purposes, and the high resolution mode is not the standard one  for  \forstwo, no  suitable  data  was  found in the archive.  Thus,  since  the November 25th science images can not be calibrated with twilight flat--fields taken on the same night, initially we do not use them in the subsequent analysis.   For  both  the \fors\  and \forstwo\  data  sets the photometric  calibration  was  applied using  the available, extinction corrected,  night zero points  available through  the instrument  data quality control   database\footnote{http://www.eso.org/observing/dfo/quality/FORS2/qc/qc1.html}. For the  January 16th and 22nd  \forstwo\ observations, for  which no standard star images were taken, we assumed as a zero point the value extrapolated from the night zero  point trend.  Since none of  the \forstwo\ observations was  taken in  perfectly photometric conditions,  we estimated  that  a relative photometry calibration,  with the February 11th observations  taken as a reference,  would  introduce  an  uncertainty comparable to  that associated to  the extrapolation  of the zero  point trend.    We converted the trended  \fors\ and \forstwo\ zero points, computed in units of electrons/s, to units  of ADU/s by applying the corresponding electrons--to--ADU conversion factors.  For  each of the two data sets,  we then used the {\em MIDAS} task {\tt average/window}  to cosmic rays filter and stack single exposures.

\begin{figure*}
\centering 
\includegraphics[height=5.9cm,angle=0,clip]{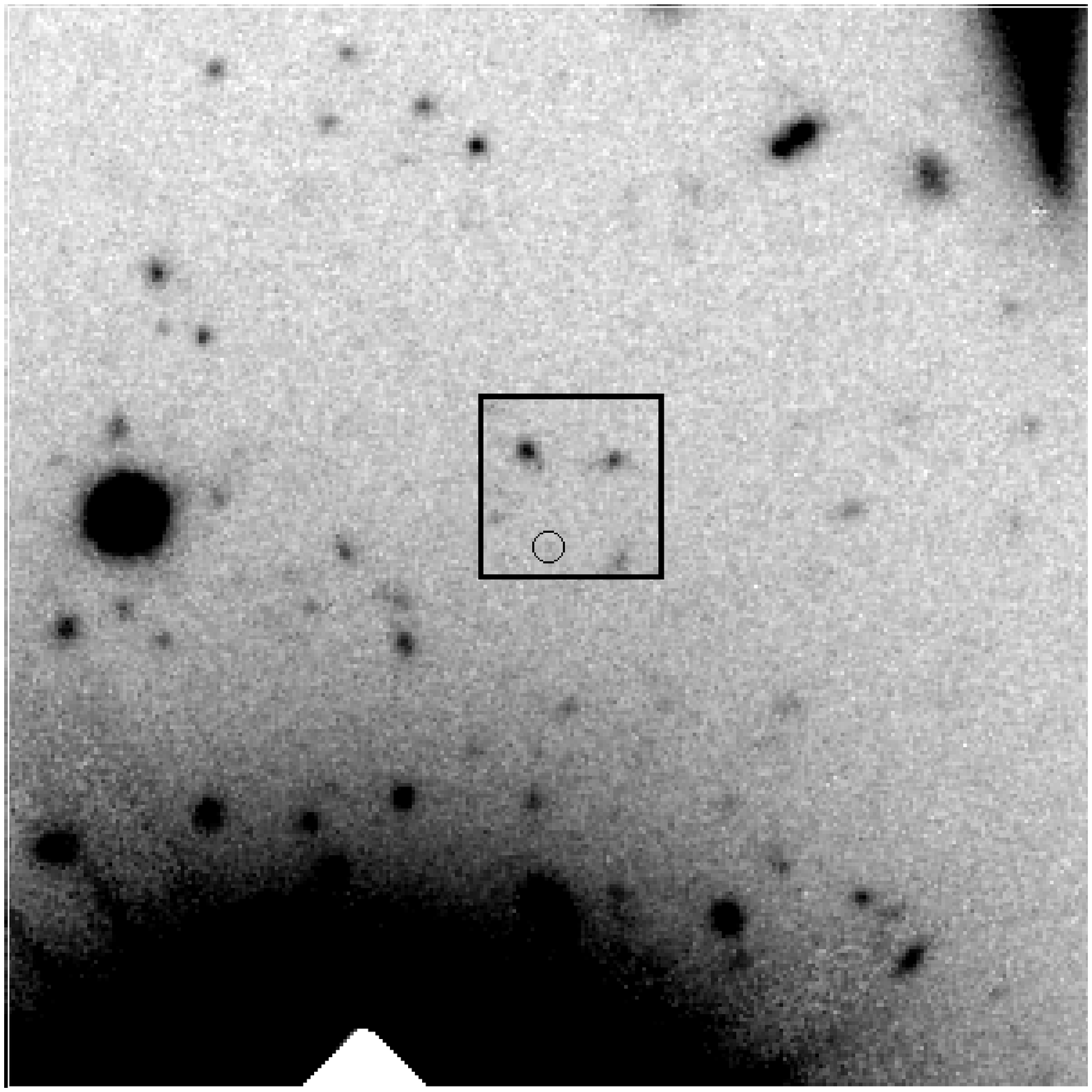} 
\includegraphics[height=5.9cm,angle=0,bb=140 130 420 410,clip]{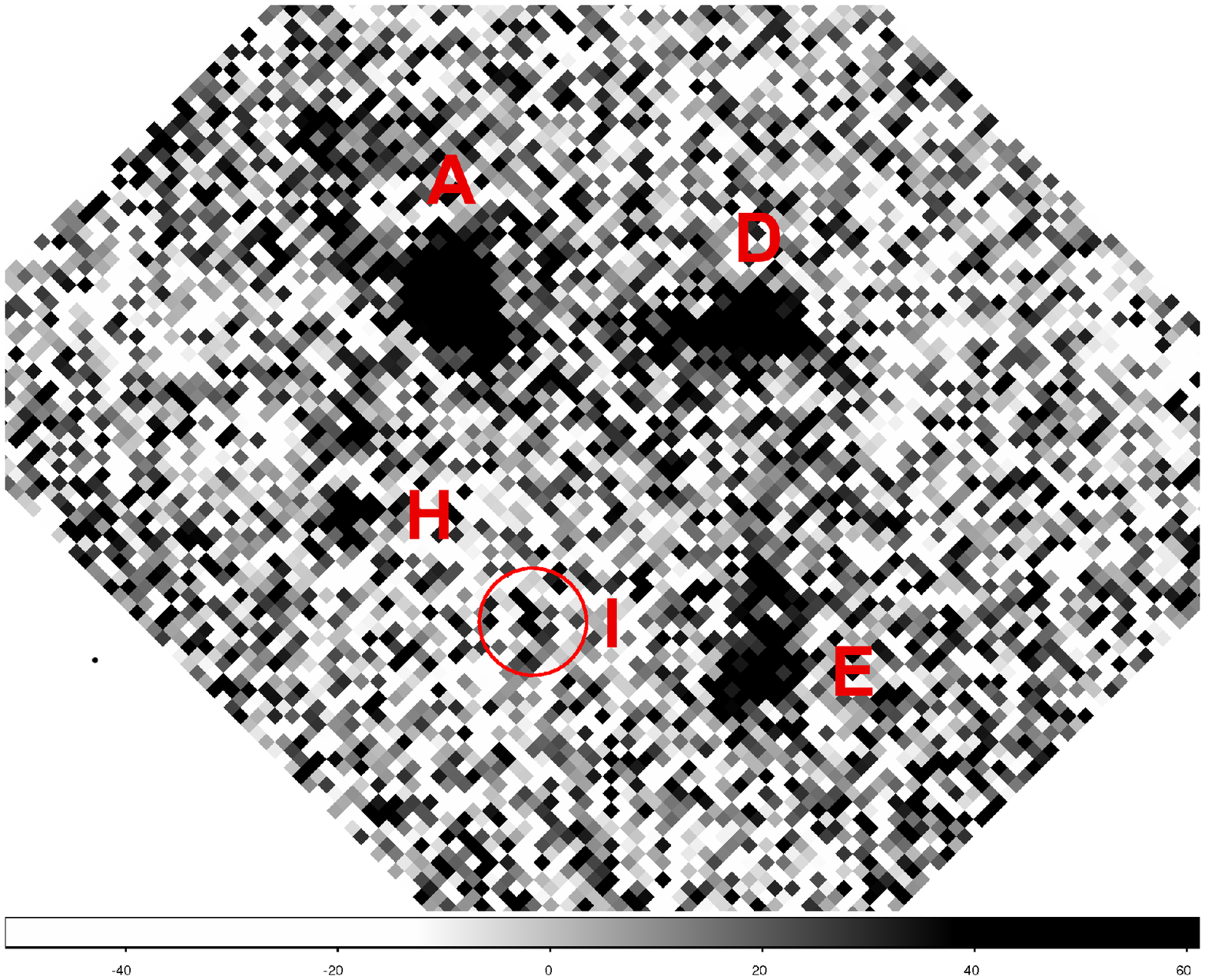} 
\includegraphics[height=5.9cm,angle=0,bb=135 95 415 375,clip]{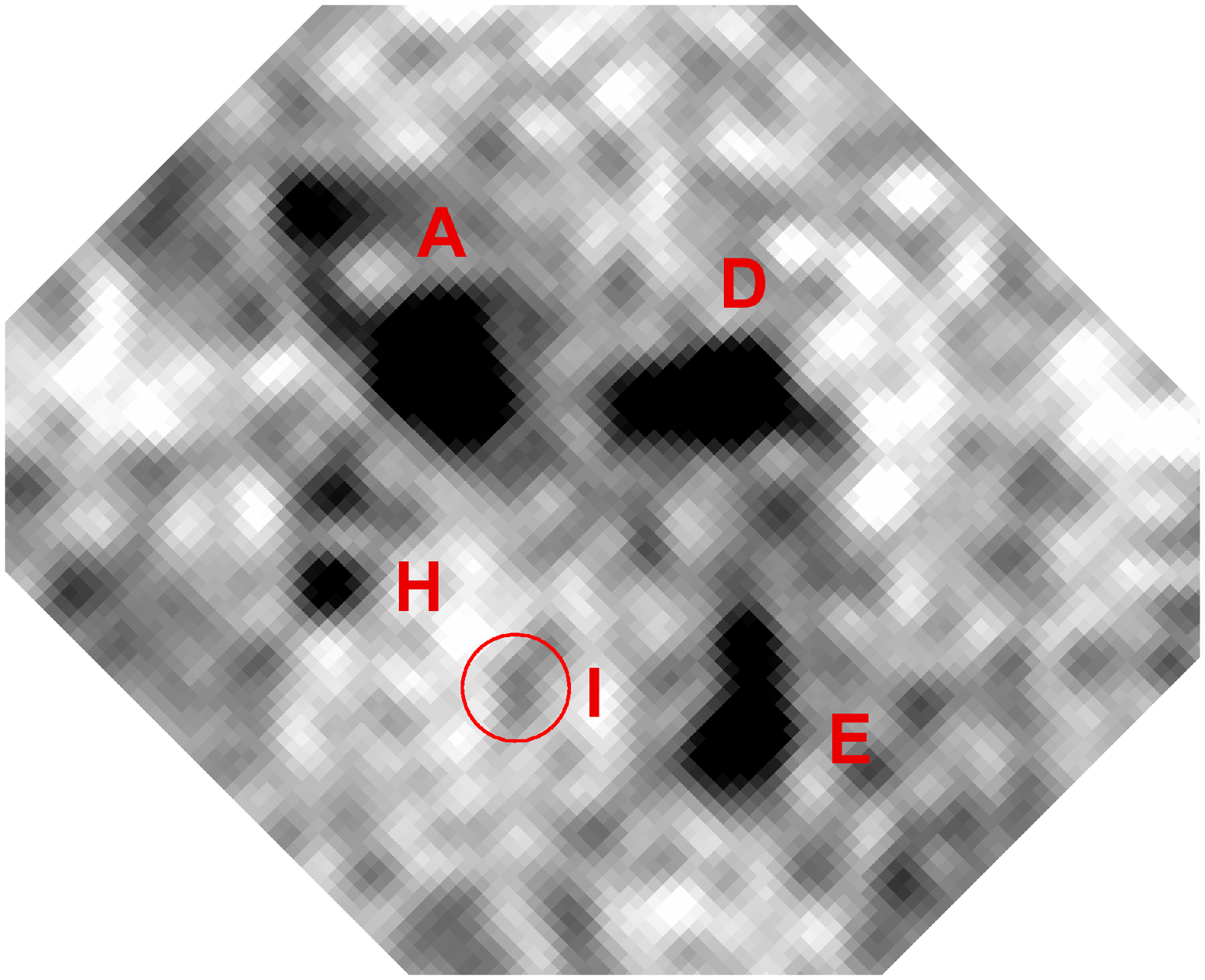}
  \caption{Left panel: $1\arcmin \times  1\arcmin$ cutout  of  the coadded \fors\ B-band  image (3600 s) of  the \zerofour\ field.   North to the top, east to the left.  The  \zerofour\ position lies at the centre of the  $10\arcsec \times  10 \arcsec$  square  and is  indicated by  the circle.   The   radius  of   the  circle  (0\farcs87; 90\% confidence level)   represents  the uncertainty on  the computed 2000 \zerofour\  position (see Sect. 2.3), which accounts for  the absolute accuracy  of the \chan\ coordinates  at the reference epoch, the uncertainty on the proper motion extrapolation at the observing epoch, and  the accuracy of our astrometric calibration. The  white triangle at  the bottom  of the  image is  the edge  of the \fors\ occulting  bar.  Middle panel: Enlargement of  the $10\arcsec \times 10 \arcsec$  region after sky background subtraction.  The intensity scale  has been adjusted  for a better visualisation of the faintest  objects in the field. Objects labelling is as in Haberl et al.  (2004).  The faint feature labelled  $I$ is their proposed candidate counterpart. Right panel:  image of  the  same  region  smoothed with a Gaussian filter over  cells of $3\times3$ pixels.  }
\label{fors1}       
\end{figure*}

\subsection{Astrometry}

As  a reference  for the  astrometric  calibration we  used the  \gsc\ version 2.3 (Lasker et al.  2008).  Approximately 70 \gsc\ objects are identified in the  2000 \fors\ image.  From this  list we filtered out extended objects, stars  that are either saturated or  too faint to be used as reliable  astrometric calibrators or too close  to the CCD edges.   We finally  performed  our astrometric  calibration using  30 well-suited \gsc\ reference stars, evenly distributed in the \fors\ field of view. The pixel  coordinates of  the selected \gsc\  stars were  measured by fitting their  intensity profiles with  a Gaussian function  using the dedicated  tool of the  {\em Graphical  Astronomy and  Image Analysis} ({\em                                                            GAIA}) interface\footnote{star-www.dur.ac.uk/~pdraper/gaia/gaia.html}.     The coordinate  transformation  between  the  detector and  the  celestial reference  frame was then  computed using  the {\em  Starlink} package {\tt ASTROM}\footnote{http://star-www.rl.ac.uk/Software/software.htm} using higher order  polynomials to accounts for the CCD distortions. The rms  of the astrometric  solution turned out to be $\approx$  0\farcs2, accounting for the rms of the fit in the right ascension and declination components.  Following  Lattanzi  et al. (1997), we estimated the  overall uncertainty of our astrometry by adding in quadrature the rms  of the astrometric fit and the precision with which  we can  register our field  on the \gsc\  reference frame. This  is estimated  as $\sqrt  3 \times  \sigma_{GSC} /  \sqrt N_{s}$, where the  $\sqrt 3$ term  accounts for the free  parameters (x-scale, y-scale, and rotation angle) in the astrometric fit, $\sigma_{GSC}$ is the    mean    positional  error    of    the    \gsc\    coordinates (0\farcs3, Lasker et al. 2008) and $N_{s}$ is  the number of stars used for the astrometric  calibration. The  uncertainty on the  reference stars centroids  is below  0\farcs01 and  was neglected.   We also  added in quadrature the  0\farcs15 uncertainty (Lasker et al. 2008)  on the tie  of the \gsc\ to  the International  Celestial Reference Frame  (ICRF).  Thus, the   overall  accuracy   of  the   \fors\  astrometry   is  0\farcs27 ($1\sigma$).    The  astrometric  calibration   of  the   \forstwo\  image was  computed in  the same  way but  with a lower number of  reference stars  due to  the smaller field  of view  of the \forstwo\  chip. The rms  of the  astrometric fit  then turned  out to be 0\farcs36.  Again, after accounting for    systematic uncertainties (see above) the overall accuracy of  the \forstwo\ astrometry is 0\farcs43 ($1\sigma$). \\
As a  reference to compute  the \zerofour\ position  we considered X-ray coordinates derived from \chan\ observations which are closest in time to our \vlt\ observations.  In particular, for the \fors\ observations  (epoch 2000.89) we used the \chan\ coordinates (epoch 2002.86) published   in  Haberl  et  al.    (2004),  i.e.   $\alpha_{J2000}=04^h  20^m  01.95^s$,  $\delta_{J2000}=  -50^\circ  22\arcmin 48\farcs1$  which have a  nominal error  of 0\farcs6  (90\% confidence level).  For the \forstwo\ observations (epoch  2007.04) we reanalysed a more recent \chan\ observations (epoch 2005.85). Like in Haberl et al. (2004), we determined the source position with the {\em CIAO} task {\tt celldetect}  and we obtained  $\alpha_{J2000}=04^h  20^m  01.94^s$,  $\delta_{J2000}=  -50^\circ  22\arcmin 48\farcs2$ (0\farcs6; 90\% confidence level).   As shown in Haberl et al. (2004) a match between the coordinates of the X-ray sources detected in the \chan\ field with those of their possible {\em USNOB.10} counterpart did not reveal any significant systematic shift. No significant shift is found between the coordinates of the same X-ray sources between the 2002 and 2007 \chan\ observations either. Thus, no boresight correction was applied to our reference coordinates.
Recently, an upper limit  on the \zerofour\ proper motion (123 mas  yr$^{-1}$, $2 \sigma$) was  obtained  with  \chan\ (Motch  et  al.  2009).   We accounted  for the proper  motion uncertainty  when we  registered the reference \chan\ coordinates  on the \fors\  and on the \forstwo\  images.   This  yields  an additional  position uncertainty due to  the unknown proper motion of  $\sim 0\farcs123$  and $\sim 0\farcs073$ ($1 \sigma$) for each of the two images, respectively. The overall uncertainty  to be attached to the  \zerofour\ position at the  epoch  of  the  \fors\  and \forstwo\  observations  was  finally obtained by adding in quadrature  the  error on the \chan\ coordinates ($1 \sigma$), the coordinate uncertainty due to  the proper motion,  and the overall error of the astrometric  calibration. This yields to uncertainties of 0\farcs87 and  1\farcs11 (90\% confidence level) on the  \zerofour\ position on the  \fors\ and on the  \forstwo\ image, respectively.

\begin{figure*}
\centering 
\includegraphics[height=5.9cm,angle=0,clip]{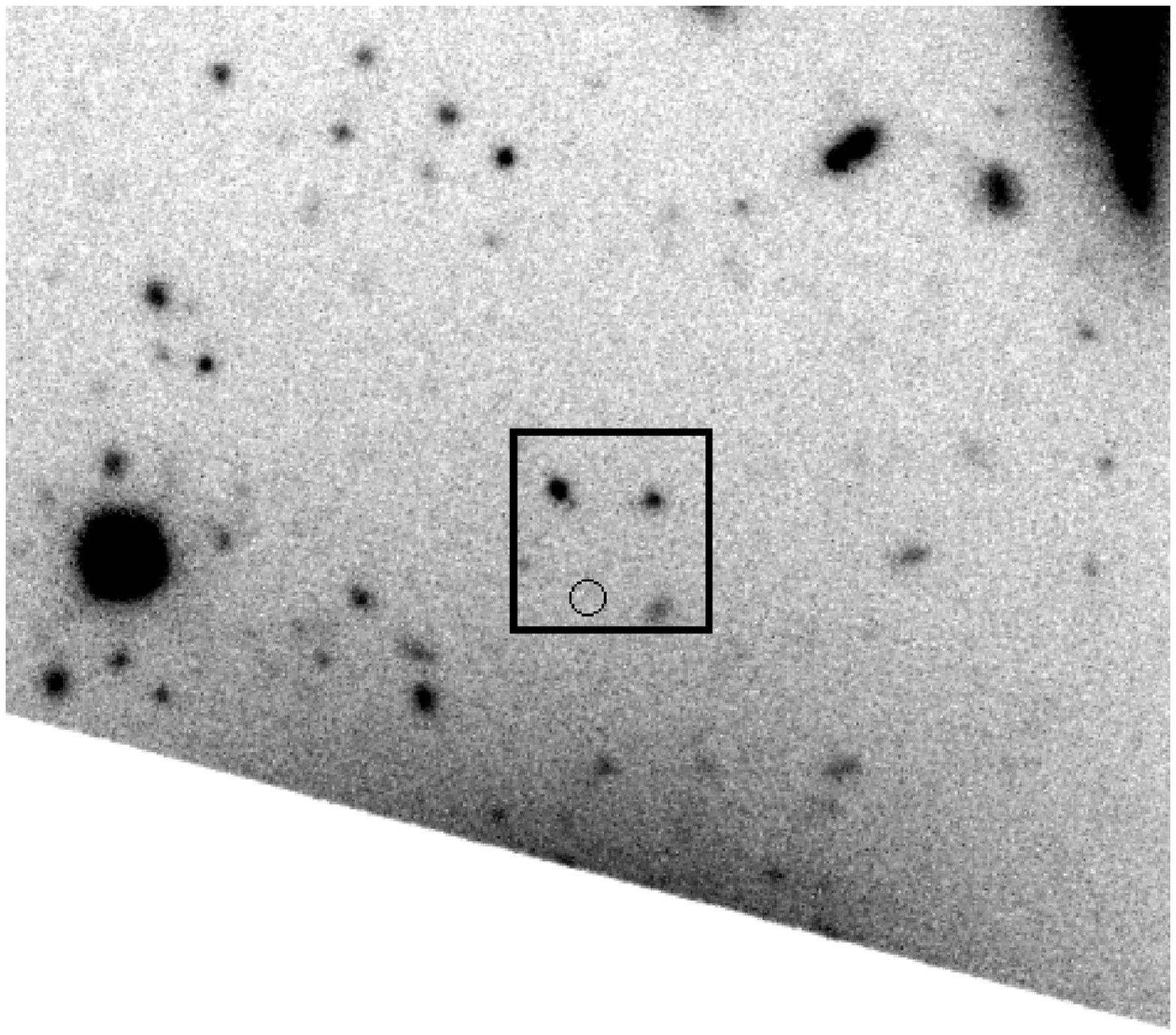}  
\includegraphics[height=5.9cm,angle=0,bb=150 100 380 330,clip]{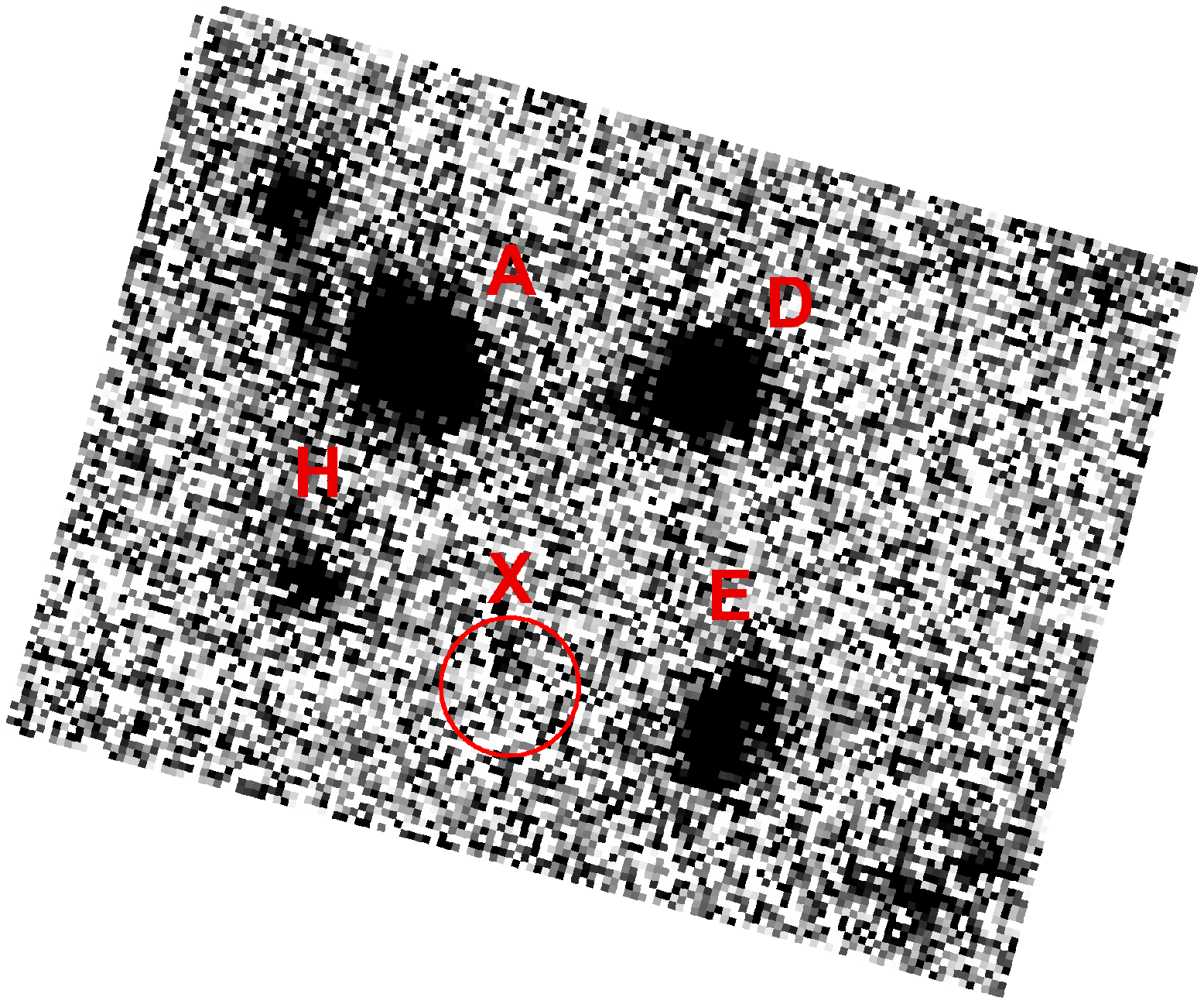} 
\includegraphics[height=5.9cm,angle=0,bb=150 100 380 330,clip]{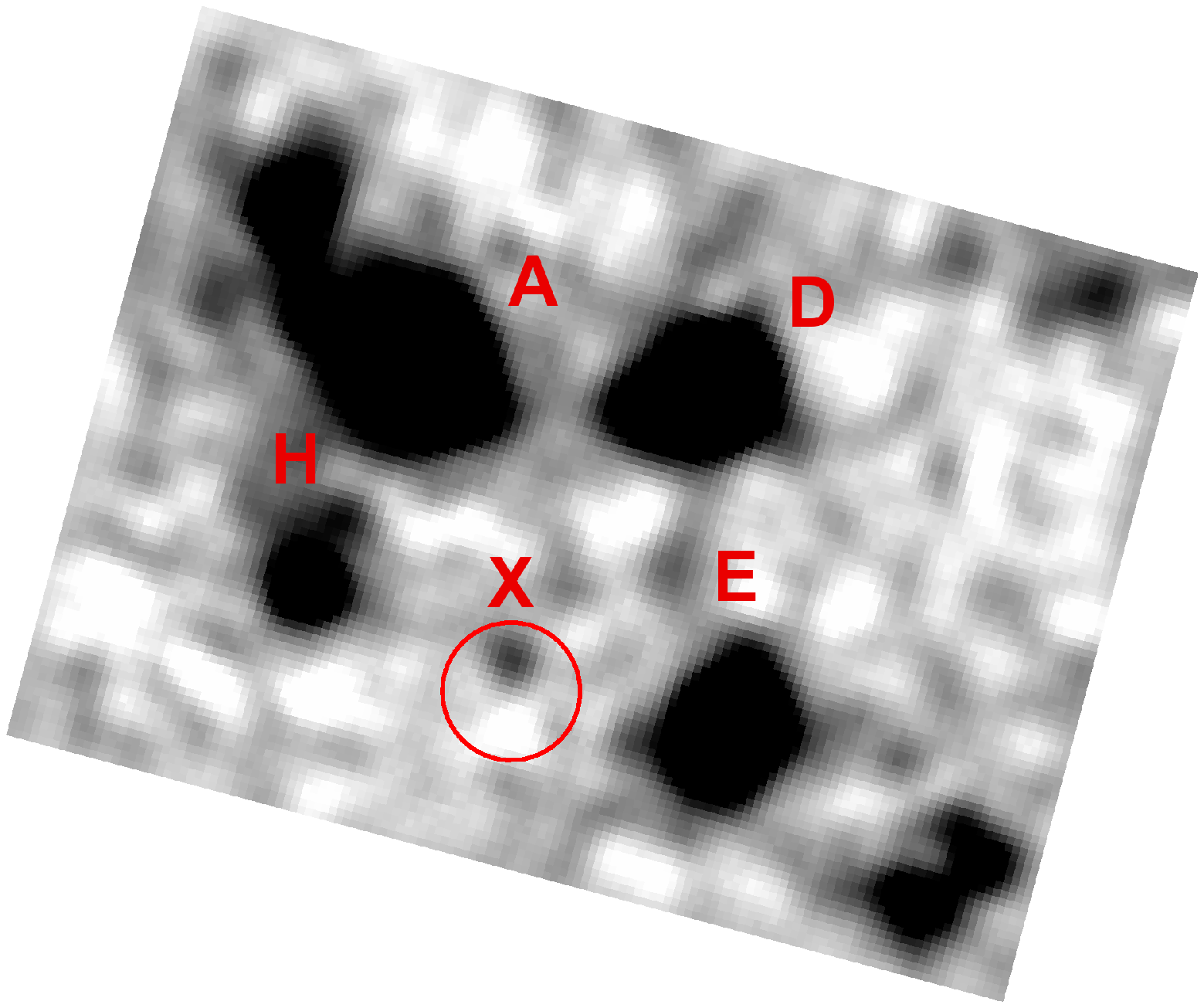}
  \caption{Left panel: $1\arcmin  \times 1\arcmin$ cutout  of the co--added   \forstwo\ B-band image  (7540 s) obtained by the  co-addition of  the January/February 2007 best  image quality exposures. North  to the top, east to the  left. The circle (1\farcs11  radius; 90 \% confidence level)  corresponds  to  the   uncertainty  on  the  computed  \zerofour\ position (see  Sect. 2.3). The white band  in the left panel corresponds to the gap between  the two CCD chips. Middle panel: zoomed image of  the inner  $10\arcsec \times 10  \arcsec$ region  (marked by  a square) after co--addition of all the available exposures (see text), rebinning, and sky background subtraction. Right panel:  image of the same region smoothed with a Gaussian filter over cells of  $5\times5$ pixels. The faint feature detected within the \chan\ error circle is labelled $X$.}
\label{fors2}       
\end{figure*}

\section{Results}

\subsection{The \fors\ observations}

We first re--analyzed the  \vlt\ observations taken in 2000 to better assess the confidence of  the optical identification of \zerofour\  proposed in Haberl et  al.   (2004).   Fig.\  1  (left) shows the  computed  \chan\  position of \zerofour\ overlaied on a cutout of the 2000  \fors\ co--added B-band  image.  As  seen from Fig.\ 1 (left), only the  bright PSF core of star  CD-50 1353 is masked, while its halo extends close to  the target  position. This increases  the  local sky  background as well as  the background noise, which results in a larger number of spurious detections.     In order to enhance the detection significance for fainter objects we tried to minimise the effects of the halo of star CD-50 1353 on the local sky background. Firstly, we fitted the sky background in an area of $\sim 15\arcsec \times 15\arcsec$ around the target position using a second order polynomial and we subtracted the fitted value from the co--added B-band image using the {\em MIDAS} task {\tt fit/flat\_sky}.  We warn here that  the fit to the sky background is biased by the choice of the sampling areas. This can yield to more or less evident feature enhancements when the sky background subtraction is applied to the image. We thus carefully  choose the sampling areas not to introduce systematic effects in our procedure. Fig.\ 1 (middle) shows a zoom of the sky-subtracted image. As already shown by  Haberl et  al. (2004), four objects are clearly  detected close to the \chan\  position.   In addition, a very faint feature is possibly recognised within the  \chan\ error circle.  We identify this feature  with object  $I$ of  Haberl et  al.  (2004),  which they tentatively proposed as a candidate counterpart to \zerofour. However, the excess of counts at the feature position is comparable to the rms of the  local sky background, which corresponds to a very low detection significance of  $\approx   2  \sigma$. We re-computed the magnitude of the feature through PSF  photometry. We derived the PSF parameters from a set of several  non saturated objects selected for their stellar like profiles, located close to the \zerofour, and spanning a large range of magnitudes. The airmass correction was applied using the Paranal extinction  coefficients measured with \fors\footnote{http://www.eso.org/observing/dfo/quality/FORS1/qc/qc1.html}. We found $B=27.52 \pm 0.61$. This is fainter than the value of $B=26.57 \pm 0.30$ reported in Haberl et al. (2004)  but it is still compatible at the $1 \sigma$ level when systematic uncertainties in their photometry are taken into account (see section 3 of Haberl et al.). For a better visualisation, we smoothed the image using a  Gaussian filter  over cells of  $3\times3$ pixels    i.e. of size  comparable to that of the  image PSF.  Since  the image smoothing  enhances the detection of very  faint objects but also that of  fluctuations of the noisy  sky background  we fine-tuned the smoothing parameters  not to produce an over-enhancement of background features. The result is shown in Fig.\ 1 (right). However,  the image processing (sky subtraction and smoothing) does not single out object $I$ against the many, similarly significant, background features recognised  around the \chan\  position.  
Thus,  we can not rule out  that object $I$ was a spurious detection due to  the high background noise induced by the  halo of star CD-50 1353.

\subsection{The \forstwo\ observations}

We used our follow-up \forstwo\  observations to search for  a higher confidence candidate counterpart to \zerofour.  In order  to minimise the effects of the halo of star CD-50 1353, we first co--added only the exposures taken with an image quality better than $1\arcsec$. In first  place, we  used the co-addition of the best image quality exposures of all nights with the exception of those taken on November 25th 2006, which were calibrated using twilight flat--fields taken about 40 days apart (see Sect 2.2).  \\
Fig.\ 2 (left) shows the computed \chan\ position of  \zerofour\  overlaid on a cutout of the  \forstwo\ B-band image (6960 s) obtained from the co-addition of the twelve best image quality (0\farcs8-0\farcs9) January/February 2007 exposures. Indeed, although the  more efficient masking reduced  the contamination from  the halo of star CD-50 1353, the  sky background  at  the \chan\  position  remained significantly affected by scattered light. As we did in Sect. 3.1, we fitted and subtracted the sky background from the co--added image.  In order to increase the S/N ratio per pixel we then rebinned the sky-subtracted image by a factor of 2, ending up with a pixel  size of 0\farcs25 which well matches   that of \fors\ (0\farcs2).  We did not find evidence for object $I$,  the feature tentatively proposed by Haberl et al. (2004) as a candidate counterpart to  \zerofour. However, we possibly recognised a second feature within the  \chan\  error circle, $\approx 0\farcs5$ north of the expected position of object $I$. Unfortunately,  the low number of counts 
only yields to a marginal detection significance ($\sim 3 \sigma$).  
As a test,  and being aware of possible issues related to the non optimal flat--fielding, we decided to use the five November 2006 exposures (2915 s) which happen to have the best image quality (0\farcs7) in the \forstwo\ data set.   As done for the January/February data set, we fitted and subtracted the sky background from the co--added image and we  rebinned the sky-subtracted image by a factor of 2. Interestingly, a feature appears right at the same position of that seen in the co--addition of the January/February best image quality exposures, although with only a  $\sim 2.5 \sigma$ detection significance.  While we do not claim that this is a strong detection evidence, it is quite unusual that a background feature appears at the same position in images taken weeks apart.  To increase the S/N ratio, we both co--added all the twenty 2007 exposures (11680 s) and all the available exposures (14575 s), again applying sky-subtraction and rebinning, and we obtained a detection significance of $\sim 3.5 \sigma$ and $\approx 3.9 \sigma$, respectively.   A zoom of the longest integration time, co--added image   is shown in Fig.\ 2 (middle), where the feature detected in the \chan\ error circle is labelled $X$.   For a better visualisation,  we smoothed the image using a Gaussian  filter  over  cells  of $5\times5$ pixels  (Fig.\ 2, right).  \\

\begin{table}
\begin{center}
  \caption{Label, coordinates, and $B$-band magnitudes of the objects identified in the \forstwo\ image (Fig.2; middle). Coordinate uncertains are derived from our astrometric calibration (Sect. 2.3).  A photometry calibration error of 0.05 magnitudes is assumed (Sect 2.2).
  }
\begin{tabular}{llll} \\ \hline
ID & $\alpha_{J2000} ^{(hms)}$ &  $\delta_{J2000}^{(\circ ~'~")}$ & $B$ 	\\  \\\hline 
 X &  04 20 01.94  & -50 22 47.75 &  27.5 $\pm$ 0.3 \\
 A &  04 20 02.10 &-50 22 42.60   &  24.35$\pm$0.05\\ 
 D &  04 20 01.59 &-50 22 43.17  &  24.95$\pm$0.05\\
 E  &  04 20 01.57 &-50 22 48.75 &  25.37$\pm$0.07\\       
 H &  04 20 02.28 &-50 22 46.30  &  26.49$\pm$0.10 \\ \hline          
\end{tabular}
\label{data}
\end{center}
\end{table}

 As done in Sect. 3.1, we  measured  the  flux of object $X$ through PSF  photometry. 
The airmass correction was applied using the Paranal extinction  coefficients measured with \forstwo\footnote{http://www.eso.org/observing/dfo/quality/FORS2/qc/qc1.html}. Due to the still low S/N in the aperture and to the noisy sky background the flux measurement is obviously affected  by a large error.  Our best estimate gives $B=27.5 \pm 0.3$, where the statistical error obviously dominates over the uncertainty of our absolute photometry  (Sect. 2.2).  The object magnitude and coordinates are listed in Table 2 together with those of the other objects identified in Fig. 2, as a reference. Interestingly, the flux of object $X$ coincides with that of object $I$ ($B=27.52 \pm 0.61$), the candidate counterpart tentatively proposed by Haberl et al. (2004), which we re-computed in Sect. 3.1.  One may thus speculate whether we  detected the same feature both in the \forstwo\ and in the  \fors\ images, although at slightly different positions. The measured angular separation between object $X$ and object $I$  is $0\farcs5 \pm 0\farcs3$,  accounting for an estimated uncertainty of one pixel on the object centroid in both the \fors\ and \forstwo\ images. This  would imply  a yearly displacement of $80 \pm 50$  mas yr$^{-1}$, consistent  with the upper limit on the \zerofour\  proper motion  (Motch  et al.   2009).  The yearly displacement would thus imply a transverse velocity of $\approx$ 140 $d_{350}$ km s$^{-1}$, where $d_{350}$  is the neutron star distance in units of 350 pc (Posselt et al. 2007), i.e. within the range of the tangential velocities inferred for neutron stars. The actual proper motion measurement, to be eventually obtained with \chan, will unambiguously address this speculation. 

\begin{figure}
\centering 
\includegraphics[height=8cm,angle=-90,clip]{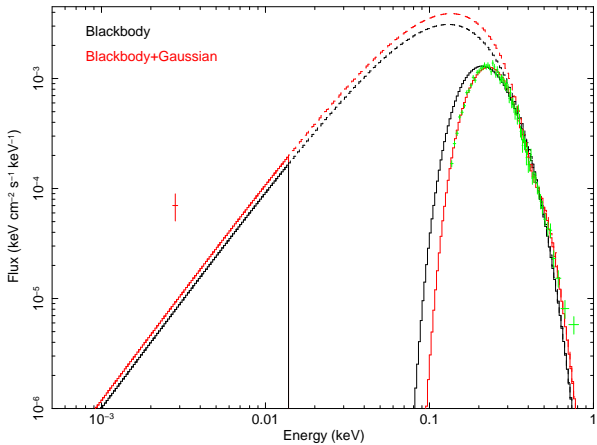} 
  \caption{Best fit models to the \xmm\ {\em EPIC-pn} spectrum of \zerofour\ (green data points). The red and black lines correspond to a single blackbody and to a blackbody plus an absorption line spectral fit, respectively (see Sect. 4 for details).  Absorption-corrected model curves are drawn as dashed lines. The dereddened B-band flux of the putative candidate counterpart ($1 \sigma$ error) is marked. }
\label{spec}       
\end{figure}

\section{Discussion}

\begin{figure}
\centering 
\includegraphics[height=6cm,angle=0,clip]{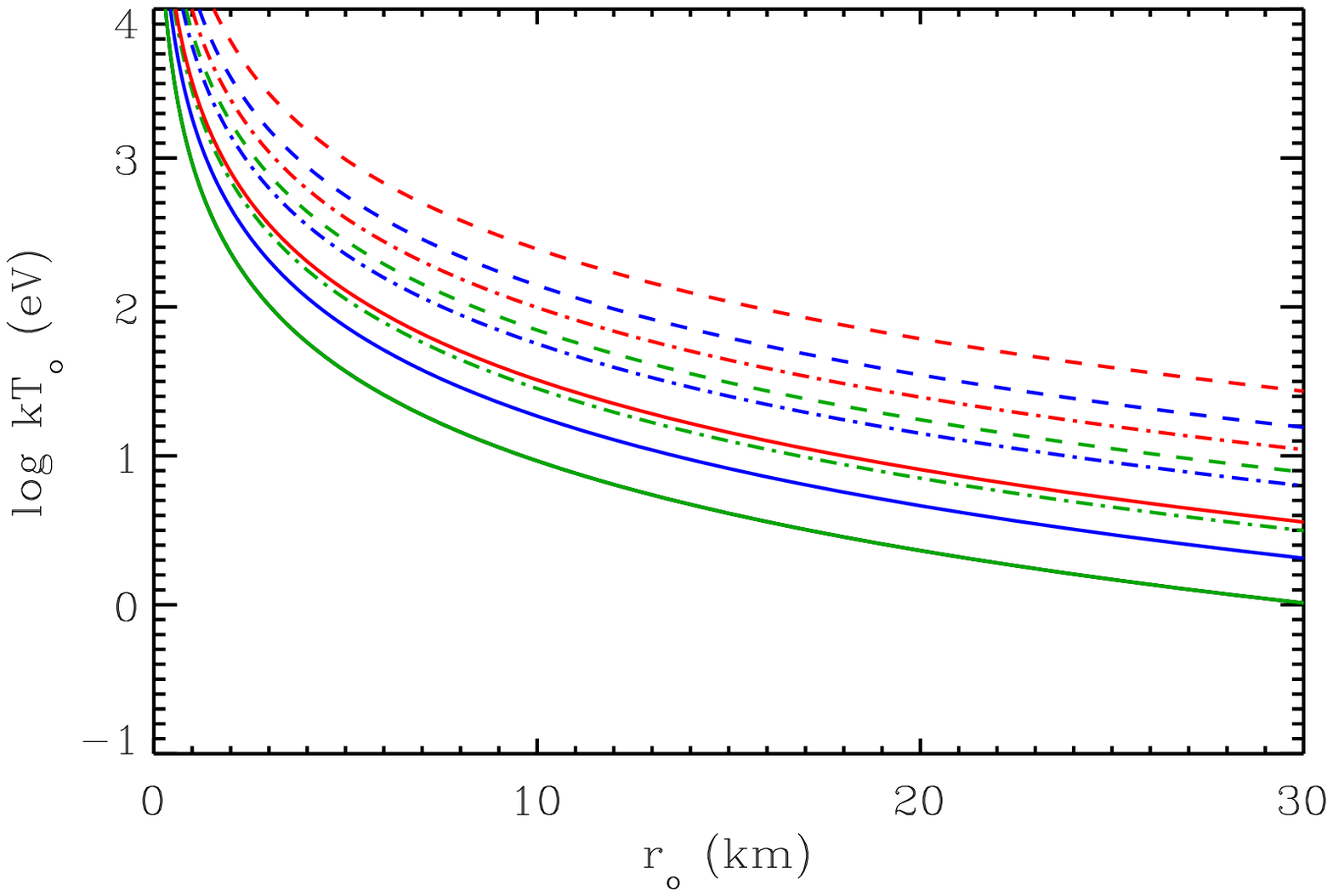} 
\includegraphics[height=6cm,angle=0,clip]{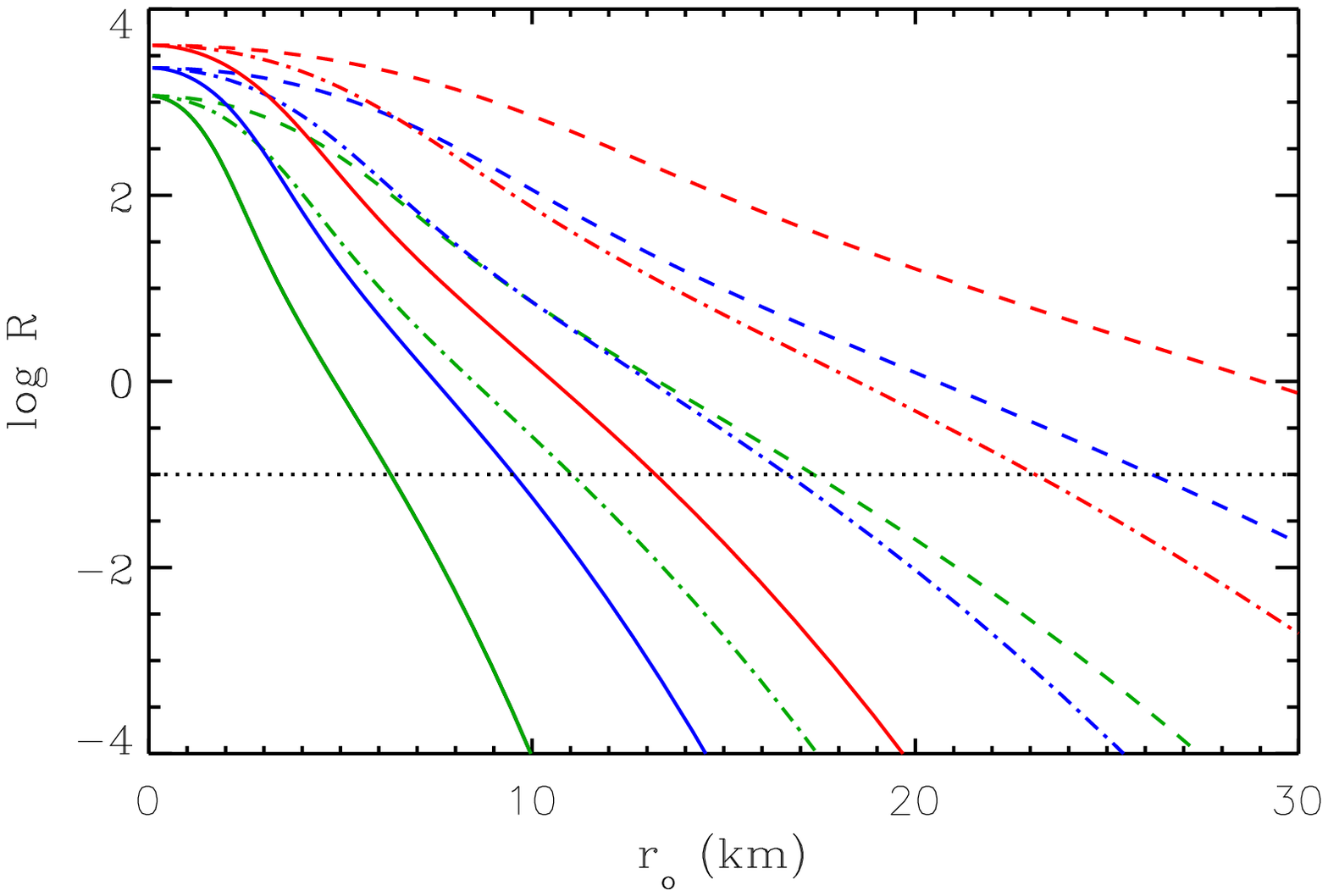} 
 \caption{
Upper panel: blackbody temperature $T_{o}$  as a function of the emission radius $r_{o}$ for different values of the optical excess $f$. The red, blue, and green  lines correspond to an optical excess of $f = 7, 4$ and 2,  respectively. For each value of $f$, the three curves are drawn for different values of the neutron star distance, 550, 350 and 200 pc (dashed, dot-dashed, and solid lines,  respectively). 
Lower panel: relative contribution $R$ to the total 0.1-1keV X-ray flux of a blackbody with temperature $T_{o}$ versus the radius of the emitting region $r_{o}$  for different values of the optical excess $f$ and of the source distance. The line style and colour coding is the same as in the upper panel. The horizontal dotted line corresponds to the threshold $R=0.1$. The allowed region in the parameters space lies below this line. }
\label{spec}       
\end{figure}

The very marginal detection significance ($\sim 3.9 \sigma$) of object $X$ against  the number of local spurious detections makes it difficult to determine whether or not it is real and, thus, whether or not we have detected a candidate optical counterpart to \zerofour.  We compared the flux of the putative candidate counterpart with the extrapolation in the optical domain of the models which best fit the \xmm\ {\em EPIC-pn} spectrum of \zerofour. To this aim, we  have re-analysed the original data of Haberl et al. (2004) using updated calibration files. The spectrum can be fit by a single blackbody with temperature $kT_{X} = 46.2 \pm 1.4$ eV and N$_{\rm H}$=$(0.73 \pm 0.21) \times 10^{20}$ cm$^{-2}$ (reduced $\chi^{2}=2.08$, 64 d.o.f.), corresponding to an emission radius $r_{X} = 5.11 ~ d_{350}$ ~ km, where $r_{X}$ is the X-ray emission as seen from infinity and $d_{350}$ is the neutron star distance in units of 350 pc (Posselt et al. 2007). However,  a blackbody with $kT_{X} = 47.8  \pm 2.2$ eV and N$_{\rm H}$=$1.19^{+0.45}_{-0.31}  \times 10^{20}$ cm$^{-2}$ ($r_{X} = 5.44 ~ d_{350}$ ~ km) plus an absorption line with centroid energy $E_{line} = 337 \pm 24$ eV and equivalent width  $EW_{line} = 47 \pm 5$ eV gives a better fit  (reduced $\chi^{2}=1.33$, 62 d.o.f.). The line  width $\sigma_{line}$) was fixed at 70 eV,  as in Haberl et al. (2004). For the spectral fits we used element abundances both from Anders \& Grevesse (1989) and Wilms et al. (2000), obtaining virtually the same results.  The best-fit, absorption-corrected X-ray spectra of \zerofour\ are shown in Fig.\ 3 together with the optical flux of its putative counterpart.  We  corrected for the absorption in the B band using as a reference the N$_{\rm H}$  derived from the best-fit X-ray spectral model (blackbody plus absorption line) and applying the relation of Predehl \& Schmitt (1995) with the extinction  coefficients of Fitzpatrick (1999).  From the flux of the putative counterpart we can rule out a $> 7$ optical excess with respect to the extrapolation of the \xmm\ spectrum. We note that an optical excess of $\sim 5$  is usually observed in other optically identified XDINSs with the exception of  \onesix\ and \oneseven, where it is as large as $\approx 15$ (Motch et al. 2005) and $\approx 30-40$ (Zane et al. 2008; Schwope et al. 2009), respectively. \\
As a limit case, we checked whether an optical excess of $\sim 7$ would be compatible with either rotation-powered emission from the neutron star magnetosphere or  with thermal emission from a fraction of the neutron star surface, colder and larger than that responsible for the X-ray emission.    In the first case, the value of the X-ray period and the upper limit on the period derivative of \zerofour\  ($P=3.45$s; $\dot {P} < 92 \times 10^{-13}$s s$^{-1}$; see Haberl 2007) only yield a rotational energy loss $\dot {E} < 8.8 \times 10^{33}$ erg s$^{-1}$.  The flux of the putative counterpart would imply an optical luminosity $L_{B} \sim 1.2 \times 10^{27}$  erg s$^{-1} ~ d_{350}^2$. This would correspond to an  emission efficiency $\eta_{B}  \equiv L_{B}/\dot{E} > 1.3 \times 10^{-7}$, which could still be compatible with the values expected for $10^{6}-10^{7}$ years old neutron stars (Zharikov et al. 2006).  However, a period derivative $\dot{P}  \sim 10^{-13}$s s$^{-1}$, comparable to that of  other XDINSs, would imply a factor of 100 lower $\dot{E}$ and would make it less likely that  the optical emission is powered by the rotational energy loss.
 In the second case,  we can constrain both the blackbody temperature $T_{o}$ and  the emission radius  $r_{o}$, as seen from infinity.  Since the fit to the \xmm\ spectrum does not require the presence of a second blackbody component at lower temperature, we can impose that its relative contribution $R$ to the total X-ray flux  in the 0.1-1 keV band  (see Sect. 3 of Zane et al. 2008) must be $<<1$. We chose $R=0.1$ as a reasonable threshold. We first computed the values of $T_o$ for a grid of values of $r_{o}$ and for different values of the optical excess $f=\frac {r_{o}^{2} ~T_{o} }{r_{X}^{2} ~T_{X}}$ and of the source distance (Fig. 4, upper panel), where $T_{X} =  47.8 $~eV  and $r_{X} = 5.44 ~ d_{350}$ km are  derived from the best X-ray spectral fit (blackbody plus absoption line, see above).  We then computed $R$ from the values of $r_{o}$ and $T_{o}$ (Fig. 4, lower panel). As it is seen, for a neutron star distance of 350 pc  an optical excess of  $\sim 7$ would be compatible with a blackbody with $kT_{o} \leq 25$ eV and an  an implausibly large emitting radius of $r_{o} \geq$ 23 km.  Thus, would our putative counterpart be confirmed, an optical excess of $\sim 7$, for a neutron star distance of $\sim 350$ pc,  might rather point towards a non-thermal origin for the optical emission, as proposed for  RBS\~ 1774 (Zane et al. 2008).  Actually, as Fig. 4 shows, unphysical large radii are required  even if the actual counterpart is dimmer, $f\sim 2$--4, unless the neutron star is at $\la 200$ pc.

\section{Conclusions}

We carefully re-analysed archival \vlt/\fors\ observations of the field of the XDINS \zerofour, taken in 2000, and we performed deeper follow-up observations with \forstwo\  in 2006 and in 2007.  With a  measured detection significance of $\sim 2 \sigma$ and a re-computed flux of $B=27.52 \pm 0.61$,  we can not rule out that  the candidate counterpart tentatively detected in the \fors\ images by Haberl et al. (2004) was  a feature of the noisy sky background, produced by the very bright nearby star CD-50 1353.  
While we could not confirm this detection in our deeper \forstwo\ images, we detected an apparently new feature ($B=27.5 \pm 0.3$) within the  updated \chan\ error circle of \zerofour,   $\approx 0\farcs5$ north of the expected position of that detected in the \fors\ images.  Interestingly, both their similar flux and their angular separation, compatible with the upper limit on the \zerofour\ proper motion, suggest that we might have actually detected the same feature both in the \fors\ and in the \forstwo\ images. However, its  still marginal detection significance ($\sim 3.9 \sigma$) makes it difficult to determine whether the latter feature is associated with a real object,  and thus it is the \zerofour\ candidate counterpart,  or it is also a possible background feature.   From the flux of the putative counterpart we can rule out a $> 7$ optical excess with respect to the extrapolation of the \xmm\ spectrum.  An optical excess of $\sim 7$ (or lower) could be compatible either with rotation-powered emission from the neutron star magnetosphere  or with thermal emission from the neutron star surface for a distance $\la 200$ pc , i.e. much lower than the current best estimate of $\sim$ 350 pc (Posselt et al. 2007).  \\
More observations are required to confirm the detection of the putative candidate counterpart and to validate its identification with \zerofour. Unfortunately, the presence of star CD-50 1353 severely hampers  ground-based follow-up observations, even if performed under sub-arcsec seeing conditions and using  a very careful masking.  High spatial resolution observations with the refurbished \hstn\ (\hst), possibly to be performed in the ultraviolet, are the only way to settle the identification issue. 

\begin{acknowledgements}
RPM acknowledges STFC for support through a Rolling Grant and thanks S. Moheler (ESO) for reducing our observations through the ESO data reduction pipeline and C. Izzo (ESO) for  technical details. SZ acknowledges STFC for support through an Advanced Fellowship. We thank the anonymous referee for his/her comments to the manuscript.
\end{acknowledgements}

\end{document}